\documentstyle[12pt]{article}
\begin{document}
\thispagestyle{empty}
\begin{center}{\Large{Early universe dynamos from  neutrino oscillations induced by torsion and pre-recombination epoch}}
\end{center}
\vspace{1.0cm}
\begin{center}
{\large By L.C. Garcia de Andrade\footnote{Departamento de
F\'{\i}sica Te\'{o}rica - IF - UERJ - Rua S\~{a}o Francisco Xavier
524, Rio de Janeiro, RJ, Maracan\~{a}, CEP:20550.
e-mail:garcia@dft.if.uerj.br}}
\end{center}
\begin{abstract}
Earlier de Sabbata and Gasperini have shown that neutrinos oscillation which gives them a mass can be induced by torsion. More recently Enqvist et al have shown that it is possible to use massive neutrinos BBN magnetic fields to seed galactic magnetic fields. Thus based on these previous investigations we present several examples of how obtaining cosmological magnetic seed fields as galactic magnetic fields from massive neutrino densities and also from the torsion obtained by Nitsch as $T\approx{10^{-24}s^{-1}}$ at the present day which yields magnetic seed field of the order of $B_{seed}\approx{10^{-12}G}$. In the case we use torsion derived from massive neutrinos given by $T_{\nu}\approx{10^{-26}s^{-1}}$ one obtains in BBN time $t\approx{1s}$ with the primordial nucleosynthesis magnetic field given by $B_{BBN}\approx{10^{11}G}$ a relic magnetic field $B_{c}\approx{10^{39}G}$ which shows that the result obtained by Enqvist et al for the cosmological fields at the early universe. Galactic dynamo seed could be obtained from neutrinos at recombination. It is also shown that in the approximation of weak fields torsion can slow down the decay of magnetic fields which confirms previous results. At Planck era where the time is $t\sim{10^{-43}s}$ and $B_{Pl}\sim{10^{58}G}$ the use of formula with the strongest torsion $10^{-19}G$this yields $B_{seed}\sim{10^{-4}G}$ is a too strong field to warrant a galactic dynamo seed.\end{abstract}

\newpage
\section{Introduction}
Investigation of dynamo equation in Riemannian pseudo-spacetime of general relativity has been undertaken by Marklund et al \cite{1}. More recently the author has shown \cite{2} that a possible generalisation of magnetic dynamo equation to Riemann-Cartan spacetime endowed with torsion is obtained where torsion is present only in the electric resistivity. Therefore in higly conductive regions of the universe torsion effects could not be present. However as it is shown here these dynamo equations in spacetime with torsion are model dependent. This can be better understood  Bamba et al \cite{3} paper where the problem of the
amplification of large scale magnetic fields at $1Mpc$ without dynamo action are given by $10^{-9}G$ in the galactic intracluster without considered magnetic diffusion. Therefore though similar in results this result takes into consideration Einstein's  teleparallel gravity where only torsion scalar is used. In this paper we consider several examples considering the magnetic field seed at nucleosynthesis era, putting together the ideas of Enqvist et al \cite{4} and de Sabbata and Gasperini \cite{5} where the first paper deals with massive or oscillating neutrinos recently discovered \cite{6} to obtain cosmological magnetic fields from massive neutrinos in the primordial nucleosynthesis era which are suitable as galactic dynamo seeds in torsioned spacetime. Instead of using string torsioned dynamos \cite{7} and of Bamba et al paper scalar torsion we use axial torsion which induce neutrino oscillations. Qualitatively torsion seems to be important for large scale magnetic fields without strong gravity of de Sabbata and Sivaram \cite{8}. Several examples of magnetic dynamo efficiency are given this spacetime with axial torsion not only from massive neutrinos but also using J Nitsch \cite{9} a torsion $T_{E}\approx{10^{-24}s^{-1}}$ where E means torsion at the surface of the earth which can be taken as as actual universe torsion value. In this case the dynamo efficiency is ${\cal{A}}\approx{10^{6}}$. In the case of cosmological magnetic fields by the stronger value of the order of $T_{c}\approx{10^{-19}s^{-1}}$ in this case the dynamo efficiency is ${\cal{A}}\approx{10^{8}}$. In this case the cosmological magnetic field strength is $B_{c}\approx{10^{-14}G}$, which is well within the bounds obtained for galactic dynamo by Enqvist et al \cite{4}. From this reasoning we observe that dynamo action is proportional to torsion strength. This shall be proved analytically in the third section of this paper by using Maxwell´s equations in gauge invariant form. The third section of this paper discusses the absence of dynamo action in certain time scales with torsion and computation of cosmological magnetic fields at today epoch. In this section we show that huge magnetic fields like $10^{39}G$ are obtained in the early universe to generate  dynamo seed fields of the order of $10^{-17}G$. Pre-recombination magnetic field are obtained without dynamo action mechanism of amplification. In section 4 addresses the conclusions.
\section{Dynamo evolution equation from torsion}
To investigate cosmological magnetic fields from nonlinear effects Kobayashi et al \cite{10} they derived a set of nonlinear evolution equations that govern magnetogenesis with vorticity. In this paper we investigate the Maxwellian part of these equations and generalizes it to Riemann-Cartan spacetime making use of the frame used by them with inertial coordinates. Their frame was given by $D_{a}f=(0,{\nabla}f){\dot{S}}_{a}=(0,{\partial}_{t})$ where $a=0,1,2,3$ and $curlS_{a}=(0,{\nabla}\times\vec{S})$ where the covariant curl is given by 
\begin{equation}
curlS_{a}= {\epsilon}_{abc}D^{b}S^{c}
\label{1}
\end{equation}
where the covariant time derivative along the flow velocity $u^{a}$ is
\begin{equation}
{\dot{S}}_{a}= {{h}_{a}}^{b}u^{c}{\nabla}_{c}S_{b}
\label{2}
\end{equation}
where the orthogonal metric is
\begin{equation}
h_{ab}= {g}_{ab}+u_{a}u^{b}
\label{3}
\end{equation}
The Faraday electromagnetic tensor splitted into electric and magnetic parts reads
$F_{ab}=2u_{[a|E_{b}]}+{\epsilon}_{abc}B^{c}$, where $u^{a}u_{a}=-1$ therefore $E_{0}=B_{0}=0$. Now let us introduce torsion using the minimal coupling 
${\nabla}_{RC}={\nabla}+\vec{T}$ and 
${{\nabla}_{RC}}_{0}={\partial}_{t}+{T}_{0}$, where $T_{a}$ represents the torsion four-vector. 	 Substitution of torsion minimal coupling into Maxwell equations in covariant form yields
\begin{equation}
{D}_{a}E^{a}+T_{a}E^{a}=-2B_{a}{\omega}^{a}+{\mu}
\label{4}
\end{equation} 
Here ${\mu}=-j_{a}u^{a}$ is the charge density which we 
put to vanish in the case of neutrinos which are chargeless. Here ${\omega}^{a}$ is the vorticity which we shall disregard soon cause we are considering the use of FRW cosmology with torsion. The equation for the magnetic part yields
\begin{equation}
\dot{B^{a}}=+T_{a}E^{a}-\frac{2}{3}{\theta}B_{a}-curl{E}^{a}-{\epsilon}_{abc}\dot{u^{b}}E^{c}
\label{5}
\end{equation}
Important relations to build vectorial Maxwell s equations are given by
\begin{equation}
\dot{B^{a}}=(0,{\partial}_{t}\vec{B}+T_{0}\vec{B})
\label{6}
\end{equation}
and
\begin{equation}
curl{E_{a}}=(0,{\nabla}\times{\vec{E}}+\vec{T}\times{\vec{E}})
\label{7}
\end{equation}
and $T_{a}u^{a}=0$ cannot be obeyed since this would imply that $T_{0}$ component of torsion would vanish and as we shall see bellow the $0-component$ is exactly the neutrino mass density component proportional to torsion which is given by ${\rho}_{\nu}$. The Maxwell equations in vector form are
\begin{equation}
{\partial}_{t}{\vec{B}}=-\frac{2}{3}{\theta}\vec{B}+\vec{\omega}\times{\vec{B}}-{\nabla}\times{\vec{E}}-\vec{T}\times{\vec{E}}-\frac{2}{3}T_{0}\vec{a}
\label{8}
\end{equation}
\begin{equation}
{\nabla}.{\vec{B}}=(-\vec{T}.\vec{B}+{2}\vec{E}.\vec{\omega})=0
\label{9}
\end{equation}
where we assumed here that there is no magnetic monopole in this universe. Thus we assume the constraint
\begin{equation}
\vec{T}.\vec{E}={2}\vec{B}.\vec{\omega}
\label{10}
\end{equation}
This relation is very interesting by itself because though here we are not taken vorticity into consideration it implies that if the electric and vorticity are present and the magnetic field is proportional to the electric field, and torsion space vector is proportional to vorticity a formula similar to Blackett formula \cite{11}. Note that in classical electrodynamics in flat spacetimes without torsion the electric field is necessary and important to generate the magnetic dynamo, however here in highly conductive phases of the universe the $curl\vec{E}$ does nor need to contribute to magnetic field cause there are still terms of expansion ${\theta}$ of the universe and due to minimal coupling the torsion 0-component shall be responsible for the amplification of the magnetic field as we shall see in the next section. Here of course the universe expansion does also have a contribution of torsion by minimal coupling as ${\theta}={\nabla}_{a}u^{a}(RC)={\nabla}_{a}u^{a}+T_{0}$ where $u^{a}=(1,0)$ is the comoving coordinate system used in cosmological fluids. Now let us conclude this section with the derived dynamo equation with torsion with universe flow. This last Maxwell equation is given by
\begin{equation}
{\partial}_{t}{\vec{E}}=-\frac{2}{3}{\theta}\vec{E}+\vec{\omega}\times{\vec{E}}+{\nabla}\times{\vec{B}}+\vec{T}\times{\vec{E}}-\frac{2}{3}T_{0}\vec{a}+\vec{a}{\times}\vec{B}-\vec{J}
\label{11}
\end{equation}
Taking into account that even the acceleration of the universe is great the magnetic fields are very weak and therefore taking into account a highly conductive universe implies that the electric field is very weak and them can be disregarded as well therefore with all the assumptions the Maxwell equations of interest here are
\begin{equation}
{\partial}_{t}{\vec{B}}=-\frac{2}{3}{\theta}\vec{B}+\vec{\omega}\times{\vec{B}}\label{12}
\end{equation}
where the torsion is included in the expansion in the minimal coupling 
\begin{equation}
{\theta}(RC)={\theta}+T_{0}\label{13}
\end{equation}
where ${\theta}_{Riem}=3\frac{\dot{a}}{a}=3H_{0}$ where Hubble constant is $H_{0}\approx{10^{-21}s^{-1}}$ this is very interesting cause in this formula in general torsion fields are of the order of $T_{0}\sim{10^{-19}s^{-1}}$, this seems that we are able to consider torsion in the early universe as important as general relativistic effects.

\section{Galactic magnetic fields from prerecombination  without dynamos with torsion}
In the structure formation region of the universe we consider that can be easily show by considering the approximate vorticity free solutions of self induction equation
\begin{equation}
B\sim{T_{0}t B_{seed}}
\label{14}
\end{equation} 
To obtain this result from this formula is enough conaider the mild torsion field $T_{0}\sim{10^{-19}s^{-1}}$ and time to substitution of the generalised torsion term in the  galaxy formation and $t\sim{10^{18}s}$ and $B=B_{G}\sim{10^{-6}G}$ yields the above required result $10^{-5}G$ which seems to be not suitable as a galactic dynamo seed since is a very strong magnetic field. Now let us assume that the today magnetic field can be obtained as 
\begin{equation}
{B}_{today}\sim{10^{-24}10^{18}10^{-5}}\sim{10^{-11}G}
\label{15}
\end{equation}
Note that according Tsagas \cite{12} early universe dynamos are possible for huge magnetic fields generated in the primordial universe. In the next section we shall handle with computation of seed fields and today field of neutrino oscillations induced by torsion. In the next section we shall try to find dynamo limit seeds of the order of $10^{-35}G$ which is the minimum seed for galactic dynamos.
\section{Galactic dynamos from neutrino oscilations induced by torsion and pre-recombination}
In this section we consider the neutrino oscillations on a neutrino cosmological sea. According to de Sabbata and Sivaram the neutrino density given by ${\rho}_{\nu}=10^{39}cm^{-3}$ yields a torsion given by
\begin{equation}
{T}_{0}\sim{\frac{Gh}{c^{3}}{{\rho}_{\nu}}}\sim{10^{-27}s^{-1}} \label{16}
\end{equation}
Note that using the value for neutrinos and the pre-recombination magnetic field $B_{rec}\sim{10^{-12}G}$ one obtains $B_{seed}\sim{10^{-26}G}$ which is within galactic dynamo seed limit. In this past computation we use the rec time is $t_{rec}\sim{10^{13}s}$. Today´s magnetic field are given by $B_{today}\sim{10^{-32}G}$. Now let us note that the neutrino oscillations induced by torsion leads also to early universe strong magnetic field of the order of $10^{38}G$ as seed fields to the primordial nucleosynthesis during $1s$ of BBN. Another example of the early universe is at the Planck epoch where the time is $t_{Pl}\sim{10^{-43}s}$ and the Planck magnetic field is as large as $B_{Pl}\sim{10^{58}G}$ as shown by de Sabbata and Sivaram. The substitution of these date into the formular with the strongest torsion used in this paper yields a seed field very weak $B_{seed}\sim{10^{-4}G}$ to warrant galactic dynamo seeds. 
\section{Conclusions} Gnedin et al \cite{13} have argued that a seed field the order of $10^{-20}G$ obtained by gravitational compression could not be useful to seed galactic dynamos even in general relativistic cosmologies. In this paper we have discussion that by using the sea of neutrinos torsion in the pre-recombination one obtains a galactic dynamo seed while in the case of primordial nucleosynthesis an extremely weak torsion field would be necessary. Since such weak torsion fields are not present in the early universe there is no chance that torsion may contribute to the galactic magnetic fields having BBN magnetic fields as seeds. In spiral galaxies seed fields of the order of $10^{-17}G$ have been observed which required huge early universe magnetic fields as the ones found in the last section. Actually the huge fields in the early universe found here are stronger than the ones found by Enqvist et al \cite{14}. Finally it is important to mention that the effect of torsion is to slow down the decay of the magnetic field \cite{15} as show in GR by Barrow and Tsagas \cite{16} making use of superadiabatic amplification. In this paper we have shown also that it is possible to use torsion motivated from topological effects \cite{17} in phase transitions to investigate its effect on dynamo mechanism in the early universe as was done recently by the author with another type of topological defects the so-called QCD domain walls with torsion \cite{18} where the primordial fields found are compatible with galactic dynamo mechanism. These QCD domain walls can be connected to ${\alpha}^{2}-dynamos$. Note that here however no direct topological defects has been used cause in topological defects torsion is in general concentrated on the defect. The use of torsion to investigate galactic and intergalactic field \cite{19} has been recently investigated by the author \cite{20}.\section{Acknowledgements}
We would like to express my gratitude to Axel 
Brandenburg for helpful discussions on the problem of dynamos and
torsion. Thanks are also due to my wife Telma Mussel and daughter Maria Carolina for their patience while this work was carried out. Financial support University of State of Rio de Janeiro (UERJ) js grateful
acknowledged.

\end{document}